\documentclass[floatfix,prl,twocolumn,superscriptaddress,showpacs,citeautoscript]{revtex4}
\usepackage{latexsym,amsmath,amssymb,multirow,rotating,xspace,times}
\usepackage{dcolumn}
\usepackage{bm}
\usepackage{textcomp} 
\usepackage{graphicx}
\usepackage[usenames]{color} 
\usepackage[psamsfonts]{euscript}
\usepackage{psfrag}
\usepackage[latin1]{inputenc}
\usepackage[T1]{fontenc} 
\usepackage[english]{babel}
\usepackage[tight]{subfigure}

\newcommand{\um}{{\textmu}m\xspace}
\newcommand{\pddt}[1]{\frac{\partial #1}{\partial t}}
\newcommand{\ddt}[1]{\frac{d#1}{dt}}

 \newcommand{\fh}{49mm}
\renewcommand{\k}{\ensuremath{\langle k \rangle}\xspace}
\newcommand{\sk}{\ensuremath{\sigma_k}\xspace}

\begin{document}


\title{Granular Rayleigh-Taylor Instability: Experiments and
  Simulations}

\author{Jan Ludvig Vinningland}%
\email{janlv@fys.uio.no}%
\affiliation{Department of Physics, University of Oslo, P.0.Box 1048,
  N-0316 Oslo, Norway}%
\author{\O{}istein Johnsen} \affiliation{Department of Physics,
  University of Oslo, P.0.Box 1048, N-0316 Oslo, Norway}%
\author{Eirik G. Flekk\o{}y} \affiliation{Department of Physics,
  University of Oslo, P.0.Box 1048, N-0316 Oslo, Norway}%
\author{Renaud Toussaint} \affiliation{Institut de Physique du Globe
  de Strasbourg, CNRS, Universit\'e Louis Pasteur, 5 rue Descartes,
  67084 Strasbourg Cedex, France}%
\author{Knut J\o{}rgen M\aa{}l\o{}y} \affiliation{Department of
  Physics, University of Oslo, P.0.Box 1048, N-0316 Oslo, Norway}%

\date[]{Published in PRL {\bf 99}, 048001 (2007)}%

\begin{abstract}
  A granular instability driven by gravity is studied experimentally
  and numerically. The instability arises as grains fall in a closed
  Hele-Shaw cell where a layer of dense granular material is
  positioned above a layer of air. The initially flat front defined by
  the grains subsequently develops into a pattern of falling granular
  fingers separated by rising bubbles of air. A transient coarsening
  of the front is observed right from the start by a finger merging
  process. The coarsening is later stabilized by new fingers growing
  from the center of the rising bubbles. The structures are quantified
  by means of Fourier analysis and quantitative agreement between
  experiment and computation is shown. This analysis also reveals
  scale invariance of the flow structures under overall change of
  spatial scale.
\end{abstract}


\maketitle

Improved understanding of granular flows would be of essential
importance to a range of industrial applications, to the study of
geological pattern forming processes, and, in general, to the
theoretical description of disordered media.

As grains become smaller the effect of the interstitial fluid becomes
more important. The result is a combination of dry granular dynamics
and the hydrodynamics of the fluid. These systems give rise to a
variety of exotic and most often poorly understood phenomena such as
fluidization \cite{herrmann98:_physic_of_dry_granul_media} and bubble
instabilities \cite{gendron01:_bubbl_propag}, quicksand
and jet formation \cite{lohse04:_creat_dry_variet_of_quick}, and
sandwich structures in systems where different particle types
segregate \cite{zeilstra06:_simul_study_of_air_induc}. While the study
of dry granular media has been extensive over the past decades, the
exploration of fluid-granular systems has been of more limited scope.

\begin{figure}[!b]
  \newcommand{\sfig}[1]{\subfigure{\includegraphics[width=41mm]{#1}}}%
  \centering%
  \sfig{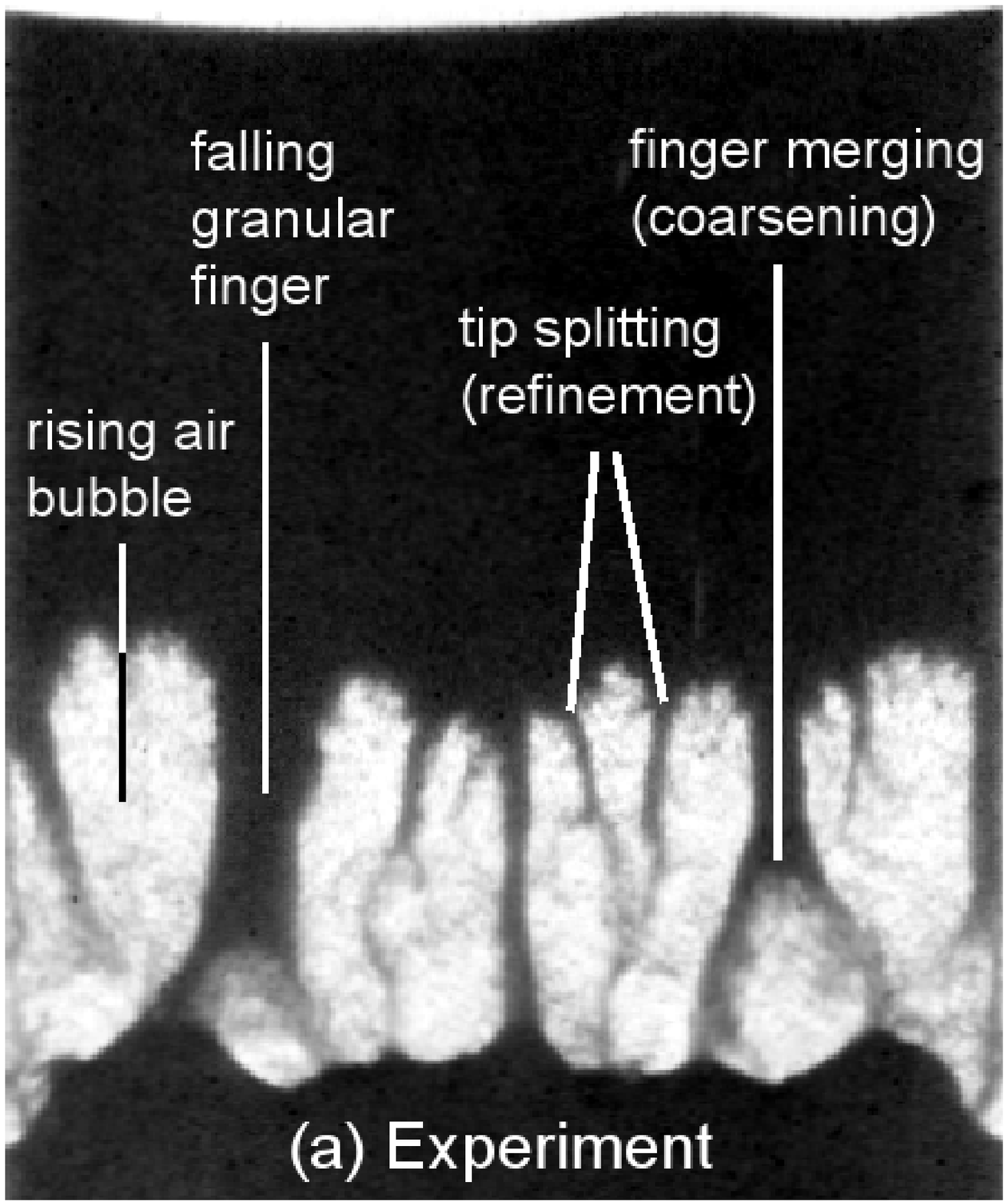}%
  \hspace{1mm}%
  \sfig{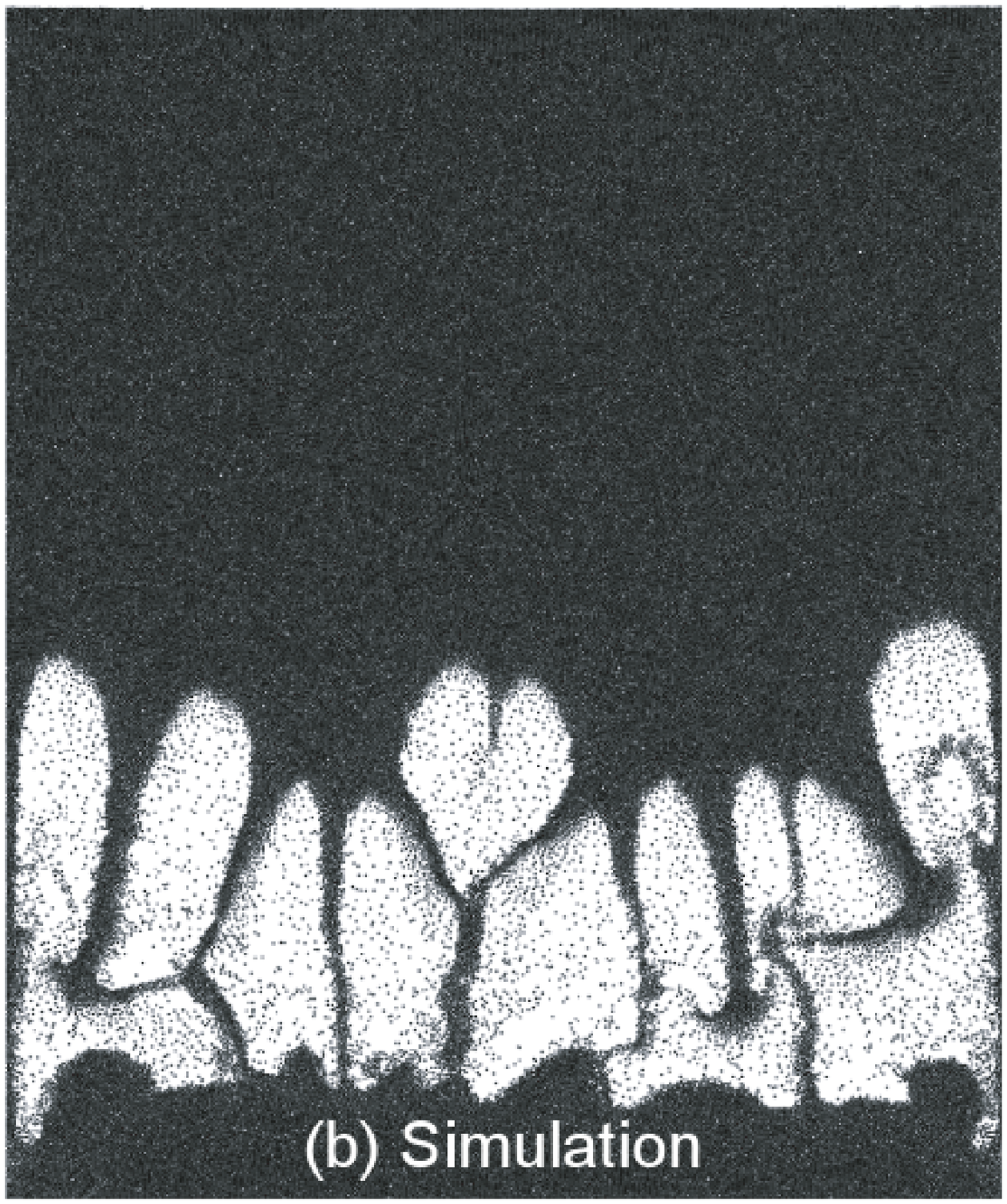}%
  \caption{(a) Experimental image and (b) numerical snapshot of a
    vertical Hele-Shaw cell where polystyrene beads (in black) of 140
    \um in diameter displace air (in white). The cells are 56 mm wide
    and were rotated 0.2 seconds ago.}%
  \label{fig:simexp}%
\end{figure}

In the present Letter we study a granular analog of the
Rayleigh-Taylor instability \cite{taylor50:instability} in the sense
that an interface instability arises as a heavier phase (the grains)
displaces a lighter phase (the air). The experimental setup consists
of a closed Hele-Shaw cell that confines air and fine grains. When the
cell is turned upside down we observe the evolution of an initially
sharp front formed by the falling grains. This evolution has three
stages: (1) An initial decompaction phase is followed by (2) the
formation of vertical falling fingers (the dark filaments in
Fig.~\ref{fig:simexp}) organizing into cusp-shaped structures that
subsequently develop into (3) coarser finger-bubble structures. The
last structures, seen in Fig.~\ref{fig:simexp}, represent a
quasisteady state where two competing mechanisms produce a
characteristic wavelength. The mechanism producing coarser scales
originates as smaller bubbles lag behind bigger bubbles, giving rise
to a finger merging process shaped like an inverted Y (see
Fig.~\ref{fig:simexp}). This process resembles the coarsening seen in
crystal growth \cite{langer80:instab_and_pattern}. The other
mechanism, that produces finer scales and is active right from the
start, is reminiscent of the tip splitting process seen in viscous
fingering. It is manifested as thin filaments forming in the centre of
the rising bubbles.

Over the past few years a wide range of granular instabilities where
various structures form along fluid-grain interfaces have been
reported
\cite{johnsen06:patter_format_durin_air_injec,gendron01:_bubbl_propag}.
Notably, the patterns formed by grains falling in a highly viscous
liquid were investigated experimentally and theoretically by V\"oltz
et al. \cite{voeltz01:_rayleig_taylor}. However, while the instability
reported by V\"oltz shares its main qualitative characteristics with
the classical Rayleigh-Taylor instability, i.e., a single dominating
wavelength growing right from the start, our gas-grain instability
grows through coarsening cusp-structures.

The evolving structures further exhibit scale invariance under change
of particle size, a feature which is supported both by observations
and theoretical considerations. The simulations and experiments that
are employed to shed light on the phenomenon at hand agree
qualitatively, and to a significant extent, quantitatively, even
though the model neglects both granular friction and the spatial
direction normal to the Hele-Shaw cell.
  
In the experiment a Hele-Shaw cell of inner dimensions 56 mm $\times$
86 mm $\times$ 1 mm is partially filled with air and monodisperse
beads (mass density 1.05 g/cm$^3$, diameter 140 \um) at atmospheric
pressure. The cell is rotated manually in about 0.2 seconds from a
lower to an upper vertical position to rapidly bring the layer of
beads above the layer of air. Images of the evolving instability are
recorded at a rate of 500 frames per second by a high speed digital
camera with a resolution of 512x512 pixels; see
Fig.~\ref{fig:simexp}(a). A simultaneous numerical snapshot is given
in Fig.~\ref{fig:simexp}(b).

The numerical model combines a continuum description of the air with a
discrete description of the granular phase \cite{mcnamara00:ggflow}.
The effect of the granular phase on the air pressure is that of a
deformable porous medium locally defined by the granular packing. The
granular phase is modeled as discrete particles from which coarse
grained solid fraction $\rho(x,y)$ and velocity fields
$\boldsymbol{\mathrm u}(x,y)$ are obtained by means of a linear
smoothing function \cite{mcnamara00:ggflow}. This function distributes
the mass and velocity of a particle among its four neighboring grid
nodes (2.5 grain diameters apart). The continuum gas phase is
described solely by its pressure $P(x,y)$. The inertia of the gas, and
hence its velocity field, is not considered. This is justified for
small particle Reynolds numbers which is the case for our system.

The pressure is governed by the equation \cite{mcnamara00:ggflow}
\begin{equation}
  \label{eq:dPdt}
  \phi \Big( \pddt{P} + \boldsymbol{\mathrm u}\cdot\boldsymbol{\mathrm \nabla}P \Big) = \boldsymbol{\mathrm \nabla}\cdot\Big(
  P\frac{\kappa (\phi )}{\mu}\boldsymbol{\mathrm \nabla}P \Big) - P\boldsymbol{\mathrm \nabla}\cdot\boldsymbol{\mathrm u}\,, 
\end{equation}
where $\phi=1-\rho $ is the porosity, $\kappa$ the permeability,
$\boldsymbol{\mathrm u}$ the granular velocity field, and $\mu$ the
gas viscosity. This equation is derived from the continuity of air and
grain mass, and Darcy's law with permeability $\kappa $. The
Carman-Kozeny relation is assumed for the permeability, and the
isothermal ideal gas law for the air.

The grains are governed by Newton's second law
\begin{equation}
  \label{eq:N2}
  m\ddt{\boldsymbol{\mathrm v}} = m\boldsymbol{\mathrm g} + \mathrm{\boldsymbol{\mathrm F}_I} - \frac{ V \boldsymbol{\mathrm \nabla}P}{\rho}\,,
\end{equation}
where $m$, $\boldsymbol{\mathrm v}$, and $V$ are respectively the
mass, velocity, and volume of the grain. Contact dynamics
\cite{radjai96:force_distr} is used to calculate the interparticle
force $\mathrm{\boldsymbol{\mathrm F}_I}$ which keeps the grains from
overlapping. The dynamics of the grains are simplified by neglecting
particle-particle and particle-wall friction.  A lower cutoff is
imposed on the solid fraction because the Carman-Kozeny relation is
not valid as the solid fraction drops below 0.25
\cite{zick82:stokes_flow}. This cutoff causes the permeability of the
most dilute regions of the system to be slightly lower than the true
permeability. The effect is a slight overestimation of the pressure
forces acting on the grains in the dilute regions.

\begin{figure*}[!t]%
  \centering%
  \subfigure[Experiment]{\includegraphics[height=\fh]{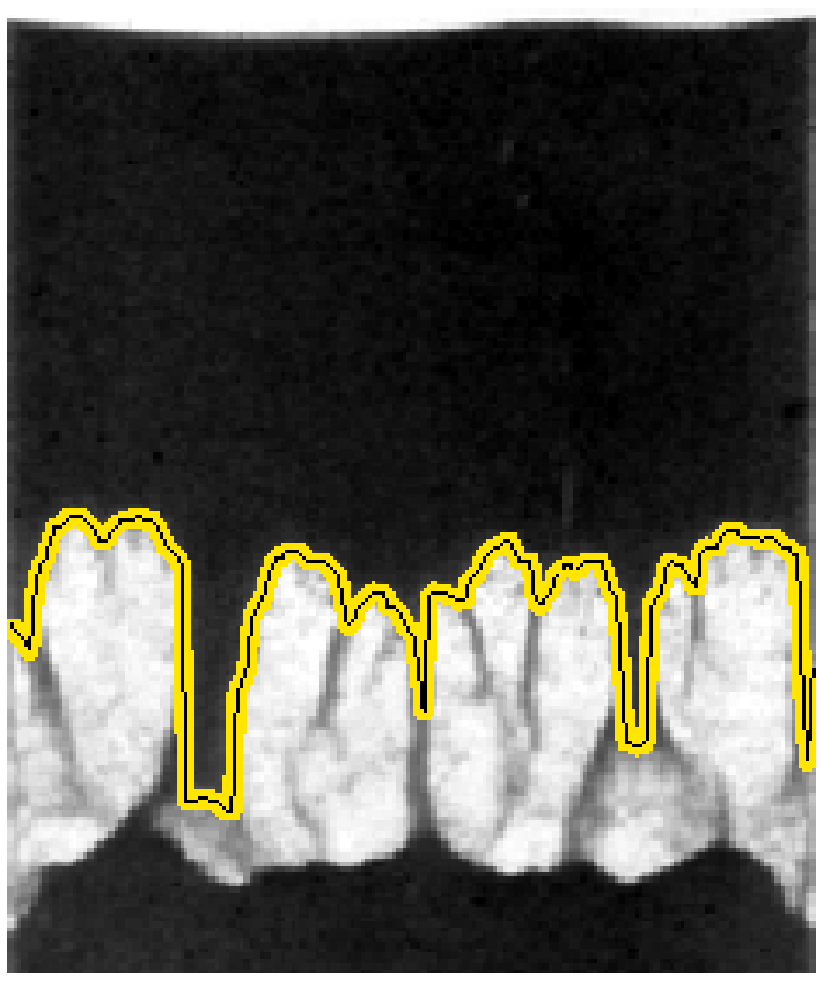}}%
  \hspace{5mm}%
  \subfigure[Experiment, interface
  evolution]{\includegraphics[height=\fh]{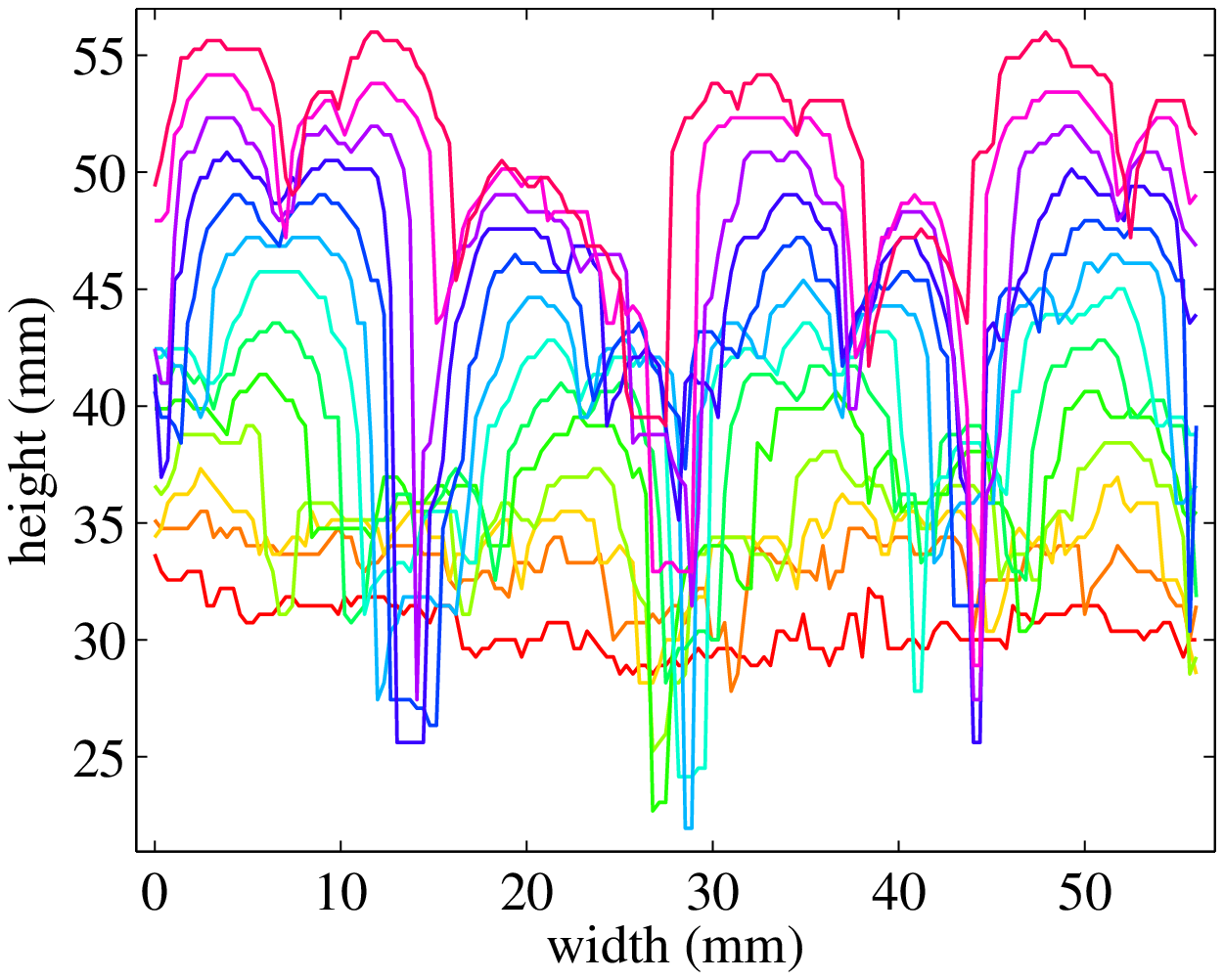}}%
  \hspace{5mm}%
  \subfigure[Experiment, interface power
  spectrum]{\includegraphics[height=\fh]{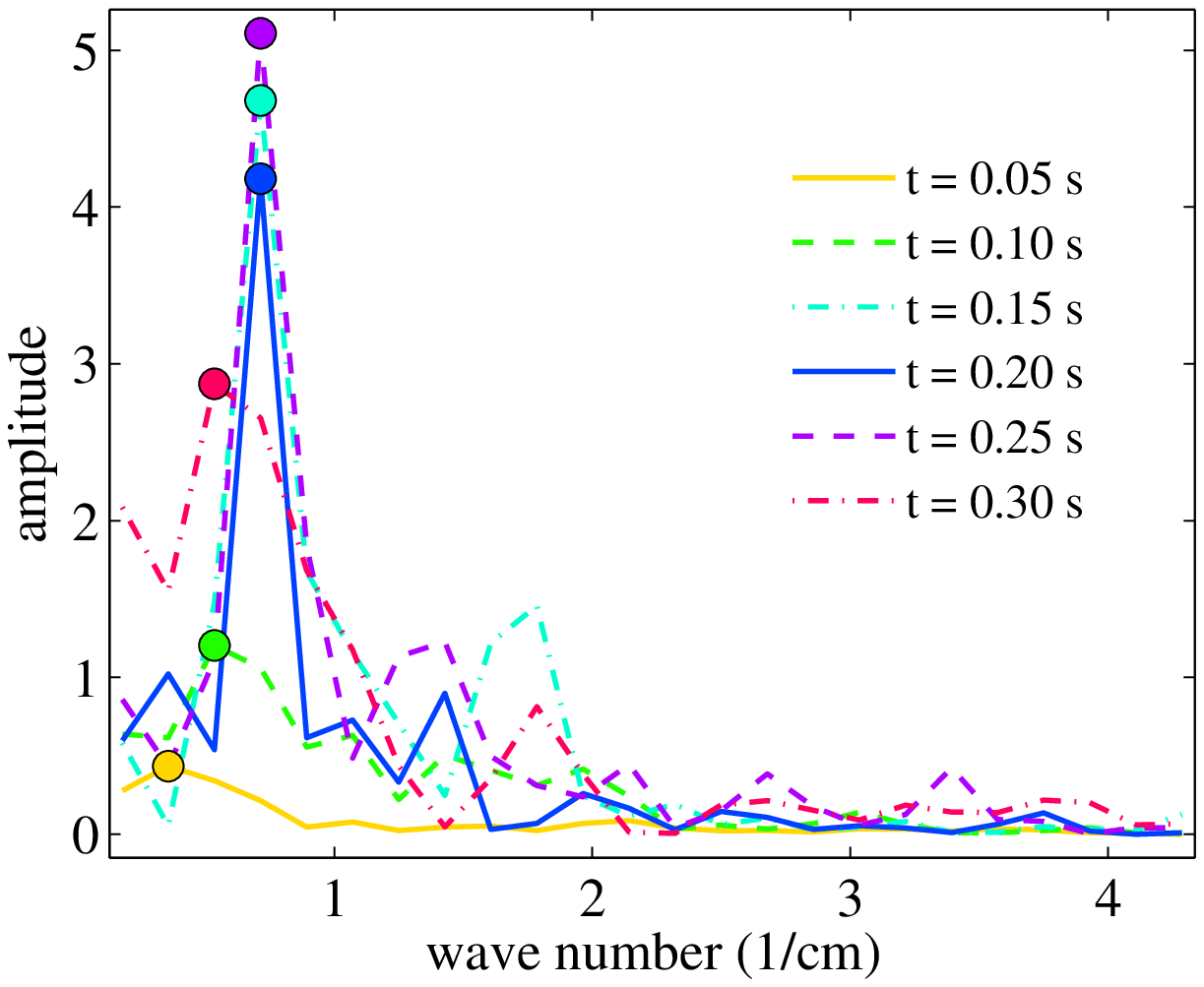}}%
  \\[2mm]%
  \subfigure[Simulation]{\includegraphics[height=\fh]{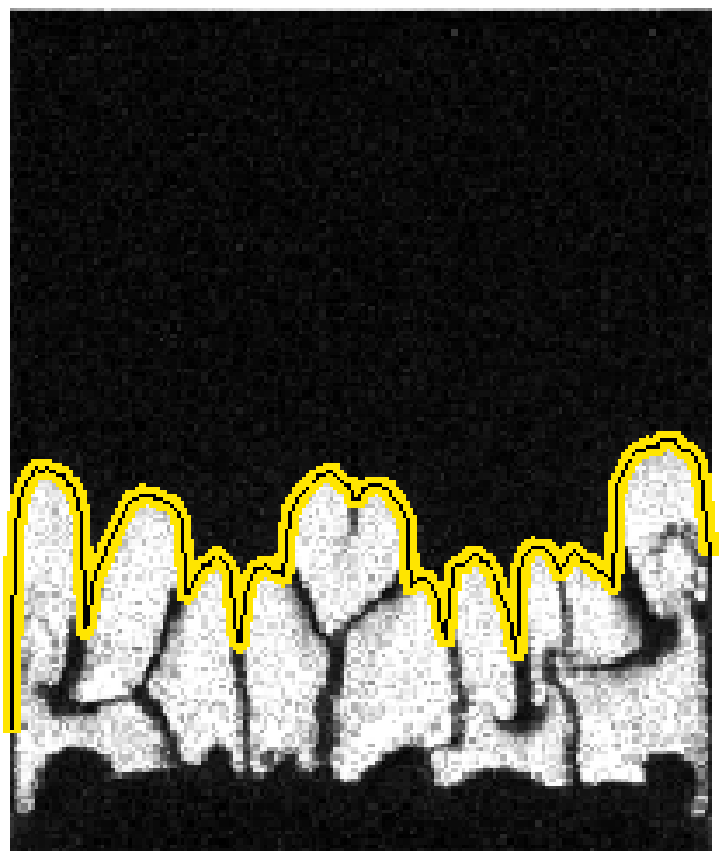}}%
  \hspace{5mm}%
  \subfigure[Simulation, interface
  evolution]{\includegraphics[height=\fh]{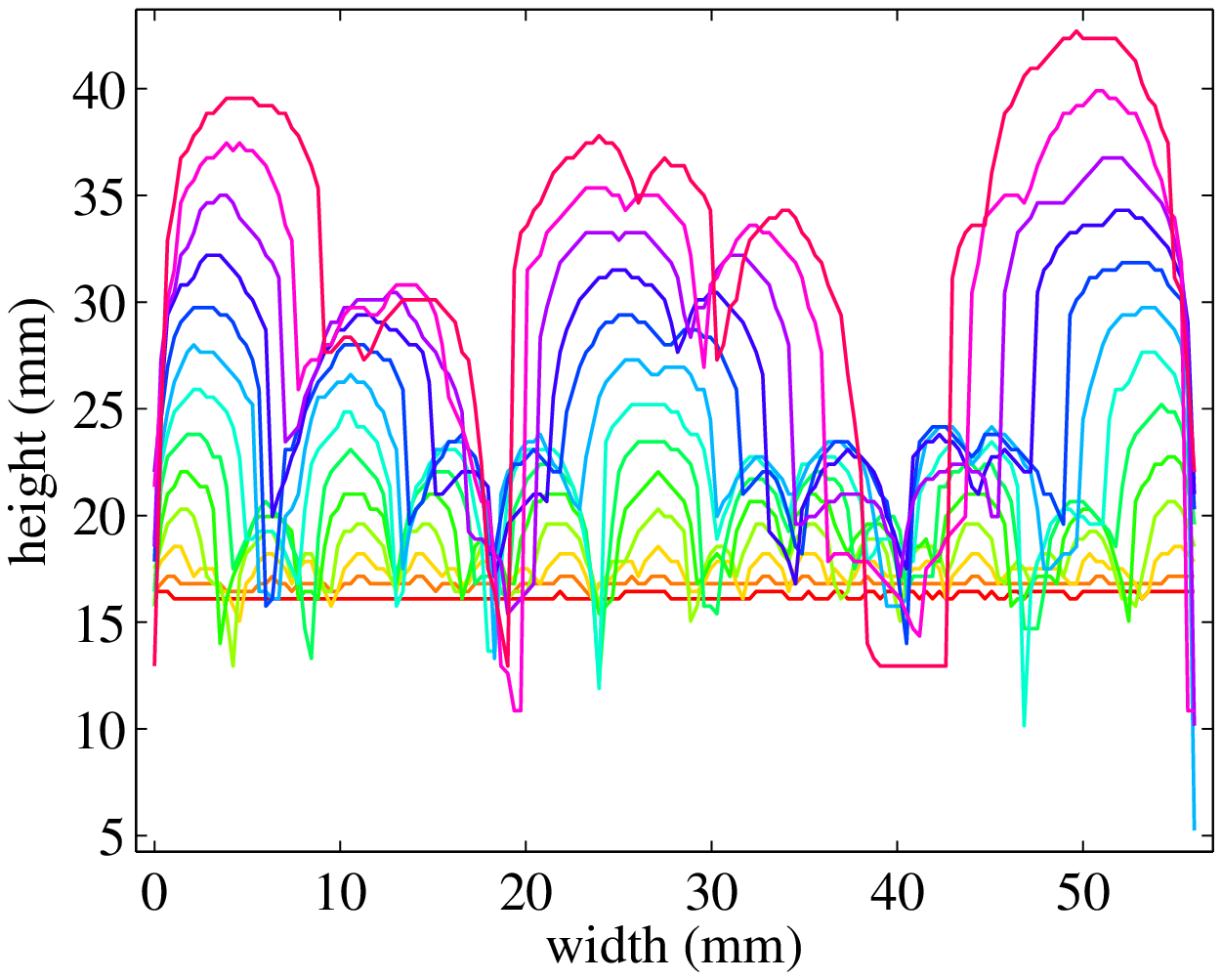}}%
  \hspace{5mm}%
  \subfigure[Simulation, interface power
  spectrum]{\includegraphics[height=\fh]{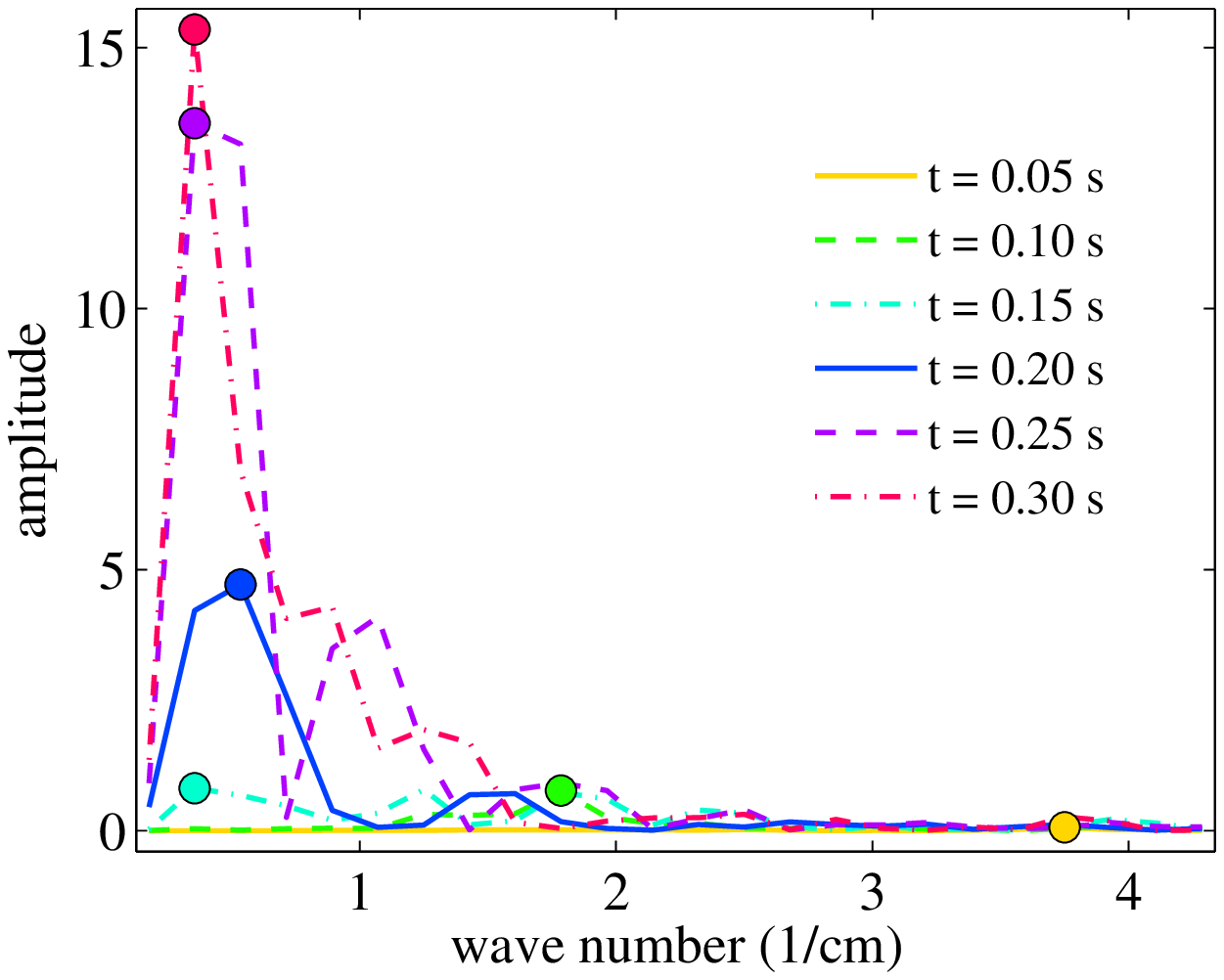}}%
  \caption{(Color online) (a),(d) Interfaces superposed on the
    snapshots ($t$ = 0.20 s) from Fig.~\ref{fig:simexp}. (b),(e)
    Spatiotemporal evolution of the air-grain interface from $t$ =
    0.00 s (bottom) to $t$ = 0.29 s (top). (c),(f) Temporal evolution
    of the power spectrum of the interfaces. The circles indicate the
    location of the maximum wave number.}
  \label{fig:interface}%
\end{figure*}

The spatiotemporal evolution of the air-grain interface in the
experiment and simulation is presented in Figs.~\ref{fig:interface}(b)
and \ref{fig:interface}(e), respectively. For every horizontal
position $x$ the interface height $y(x)$ is defined in the following
way: Moving down from the top, $y(x)$ is the height where $\rho(x,y)$
drops below a given threshold value [see Figs.~\ref{fig:interface}(a)
and \ref{fig:interface}(d)]. Because of the large density contrasts in
the system the interface position is rather insensitive to the exact
value of this threshold. For the experimental data the threshold is
set on the gray levels in the images.

The shape of the initial interfaces in Figs.~\ref{fig:interface}(b)
and \ref{fig:interface}(e) is quite different. The initial
experimental interface has noise on all wavelengths, whereas the
initial numerical interface is virtually flat with noise dominantly at
smaller wavelengths. Perturbations introduced in the cell by the
rotation and sudden stop disturb the initial experimental
interface. However, as the instability evolves the discrepancy between
experiment and simulation reduces and the later interfaces are in
better agreement.

In order to give a more quantitative comparison of the interfaces, the
discrete Fourier transform with a Hamming window is applied on every
second interface in Figs.~\ref{fig:interface}(b) and
\ref{fig:interface}(e) to produce the power spectra presented in
Figs.~\ref{fig:interface}(c) and \ref{fig:interface}(f). The power
spectra are colored as their corresponding interfaces, and the
location of the maximum wave number for each power spectrum is
indicated by a circle.  While the maximum wave number of the numerical
interfaces moves from high values to low values, the maximum wave
number of the experimental interfaces hardly moves at all, most likely
because the experiment does not evolve from an initially flat
interface. However, the experimental and numerical power spectra
converge to approximately the same form when normalized.

To study the coarsening of the observed structures quantitatively we
perform an average over the solid fraction for the entire system: The
discrete Fourier transform and the power spectrum of each horizontal
line of $\rho(x,y)$ is calculated. The averaged power spectrum,
$\bar{S}(k)$, is then obtained by averaging over all these horizontal
power spectra. An average wave number is defined as $\k = \sum_k
\bar{S}(k) \cdot k /\sum_k \bar{S} (k)$, where $1/k$ is the
wavelength. Likewise, we define the squared standard deviation
$\sk^{\phantom{k}2} = \sum_k \bar{S}(k)\cdot k^2 /\sum_k \bar{S}(k) -
\k^2$. For the experimental data the image pixel values are used to
estimate the solid fraction.

\begin{figure}[!b]%
  \centering%
  \includegraphics[width=65mm]{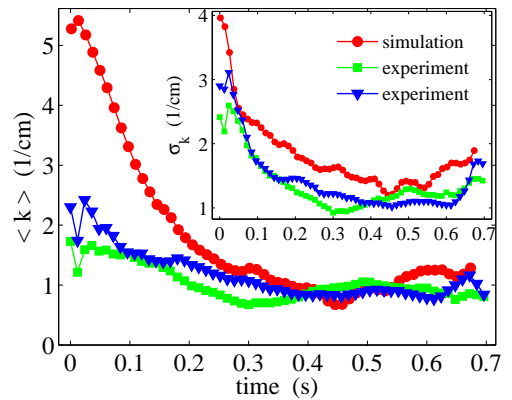}%
  \caption{(Color online) Mean wave number $\k$ and standard deviation
    $\sk$ (inset) for two experiments and one simulation, all using
    polystyrene beads of 140 \um in diameter.}%
  \label{fig:mean_std_k}%
\end{figure}

Figure~\ref{fig:mean_std_k} shows the temporal evolution of $\langle k
\rangle$ and $\sk$ (inset) for the numerical and experimental data. An
additional set of experimental data is added to the plot. The
numerical curve starts out with a significantly higher wave number
than the experimental curves. However, the numerical data decreases
monotonously until it coincides with the experimental data at about
0.2 seconds, after which the simulation and experiments show a similar
coarsening behavior. Fingers are not observed in the experiment until
0.06 seconds have elapsed. During this time the grains merely form a
dilute sheet that appears homogeneous on the experimental images. This
particular experimental initial state is caused by the sudden stop of
the cell and is the most probable reason for the initial discrepancy
between simulation and experiment in Fig.~\ref{fig:mean_std_k}. The
fluctuations of $\k$ and $\sk$ are associated with the continuous
nucleation and merging of fingers.

We further investigate the behavior of the system as the overall
spatial scale is changed: Keeping all length ratios and the particle
number fixed, the size of the system will scale according to the
diameter $d$ of the grains. We measure the characteristic inverse
length scale $\k$ as $d$ is changed and observe a scale invariance of
the evolution. A series of seven simulations are performed where $d$
varies from 70 \um to 490 \um in steps of 70 \um. The dimension of the
numerical cell confining grains of 70 \um in diameter is 28 mm x 34
mm. In these simulations we have introduced the larger density of
glass, rather than polystyrene, in order to minimize the numerical
artifacts associated with the solid fraction cutoff in the
permeability. To compare, a series of experiments using polystyrene
beads of 80, 140, 230, and 570 \um in diameter, confined in Hele-Shaw
cells that scale proportionally with $d$ in all directions, are
performed.

\begin{figure}[!t]%
  \centering%
  \subfigure[Simulation]{\includegraphics[width=65mm]{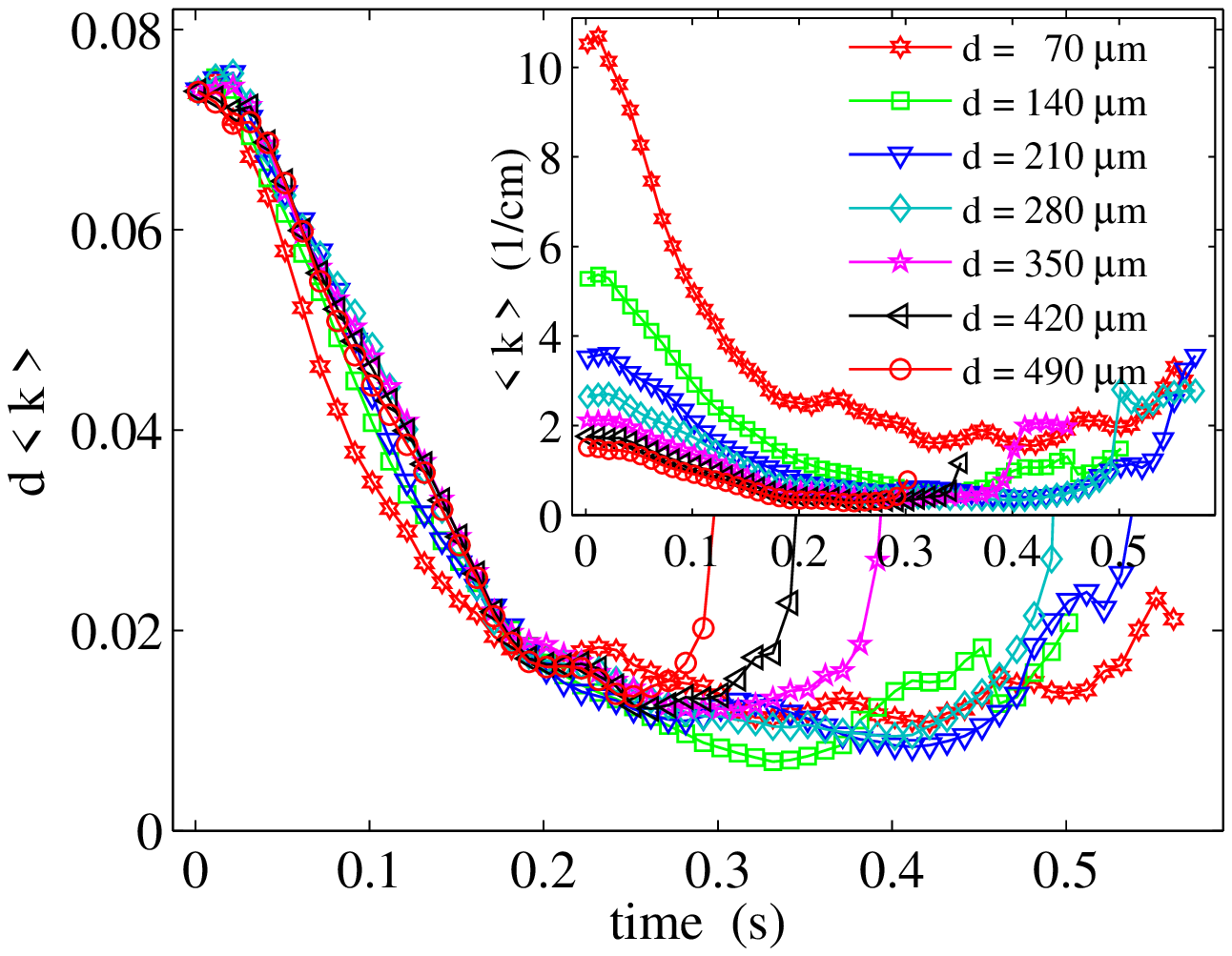}}\\%
  \subfigure[Experiment]{\includegraphics[width=65mm]{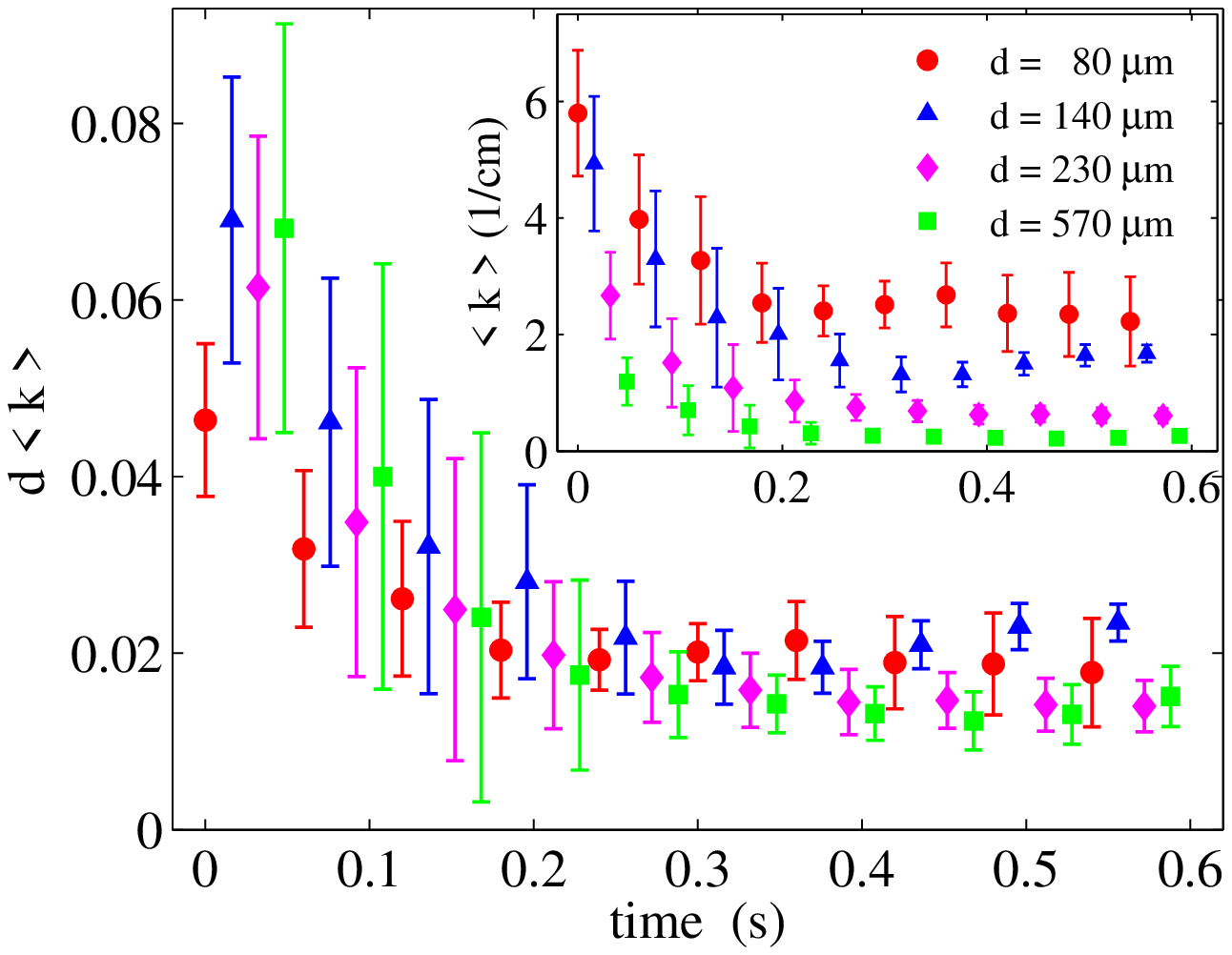}}%
  \caption{(Color online) Data-collapse plot of $d\k$ for a series of
    (a) simulations using glass beads, and (b) experiments using
    polystyrene beads. The grain diameters $d$ are given in the
    legend box. The inset shows the evolution of $\langle k
    \rangle$.}%
  \label{fig:dia_mean_k}%
\end{figure}

Data-collapse plots of the rescaled mean wave number $d\k$ are shown
in Figs.~\ref{fig:dia_mean_k}(a) (simulation) and
\ref{fig:dia_mean_k}(b) (experiment). These plots indicate that the
characteristic size of the structures is invariant when size is
measured in units of $d$; the number of grains that spans the width of
the bubbles is the same for a wide range of grain sizes.

%
%
Theoretically, the scale invariance of the product $d\k$ may be
interpreted as follows: Compared to the other terms of
Eqs.~(\ref{eq:dPdt}) and (\ref{eq:N2}) the $m\,d\boldsymbol{\mathrm
  v}/dt$, $F_{\mathrm I}$ and $P\boldsymbol{\mathrm
  \nabla}\cdot\boldsymbol{\mathrm u} $ terms may be shown to be small
\cite{vinningland06:_size_invar_in_granul_flows}. For that reason,
these equations exhibit an approximate invariance under system size
scaling. If we take $\delta P$ to be the pressure deviation from the
background pressure, express the velocity of grain $i$ as
$\boldsymbol{\mathrm v}_i = \delta \boldsymbol{\mathrm v}_i +
\boldsymbol{\mathrm u}_0$ and the locally averaged granular velocity
as $\boldsymbol{\mathrm u} = \delta \boldsymbol{\mathrm u} +
\boldsymbol{\mathrm u}_0$, where $\boldsymbol{\mathrm u}_0$ is the
constant sedimentation velocity of a close packed system, this scaling
may be expressed as $\boldsymbol{\mathrm x} \rightarrow {\lambda}
\boldsymbol{\mathrm x}$, $\delta P \rightarrow {\lambda} \delta P $,
$\boldsymbol{\mathrm u}_0 \rightarrow {\lambda^2} \boldsymbol{\mathrm
  u}_0 $, $\delta \boldsymbol{\mathrm u} \rightarrow {\lambda} \delta
\boldsymbol{\mathrm u} $ and $\kappa \rightarrow \lambda^2 \kappa$,
where $\lambda $ is a scale factor. The structure formation of the
system is governed by $\delta \boldsymbol{\mathrm u}$ and, since this
velocity scales the same way with $\lambda$ as the length scales
themselves, the evolution of any structure measured in units of $d$
will be scale invariant. In particular this is true for the structures
measured by the length $1/\langle k \rangle$, and so $d\langle k
\rangle$ is scale invariant. However, the invariance deteriorates both
when particle size is increased, and when it is decreased. In the
first case, the relative effect of granular inertia is increased, in
the second, the relative effect of the $P \boldsymbol{\mathrm \nabla}
\cdot \boldsymbol{\mathrm u} $ term is increased.

The convergence of the numerical data-collapse in
Fig.~\ref{fig:dia_mean_k}(a) is quite good. The deviation of the 70
\um curve for small $t$ is probably explained by the increase in the
relative importance of the $P \boldsymbol{\mathrm \nabla} \cdot
\boldsymbol{\mathrm u} $-term. The divergences of the 350, 420, and
490 \um curves for greater $t$ in the same plot arise because the
bubbles in the coarser packings disappear before they reach the
surface due to the increase of $\boldsymbol{\mathrm u}_0$ with
$\lambda^2$ \cite{vinningland06:_size_invar_in_granul_flows}. The
experimental data in Fig.~\ref{fig:dia_mean_k}(b) have a wider
distribution but collapses satisfactorily given the standard deviation
error bars. The experimental data are obtained by averaging over three
experiments for each diameter $d$. The standard deviation is
calculated over a time window of 0.3 seconds. The accuracy of the
experiments is at its lowest during the initial coarsening of the
structures, but as the mean wave number stabilizes around 0.2 seconds
the accuracy improves. Nevertheless, the data points are, with a few
exceptions, within a distance of one standard deviation from one
another. The loss of precision for small times is most likely caused
by the inaccuracy involved with the manual rotation.

In conclusion, we have presented experimental and numerical results of
a gravity-driven granular flow instability which is significantly
different from its classical hydrodynamic analog. The simulations
reproduce the characteristic shape and size of the experimentally
observed structures and provide fine patterns in the early phase of
the process that are not resolved experimentally. Data-collapse plots
of the mean wave number \k indicate that the flow and the resulting
structures are invariant when measured on a scale proportional to the
grain diameter $d$ for a range of diameters that spans from 70 \um to
570 \um.

\end{document}